\documentclass{article}

\usepackage[final]{nips_2018}

\usepackage[utf8]{inputenc} 
\usepackage[T1]{fontenc}    
\usepackage{hyperref}       
\usepackage{url}            
\usepackage{booktabs}       
\usepackage{amsfonts}       
\usepackage{nicefrac}       
\usepackage{microtype}      

\setcitestyle{numbers}
\setcitestyle{square}
\setcitestyle{comma}
\usepackage{graphicx}

\title{The Importance of Socio-Cultural Differences for Annotating and Detecting the Affective States of Students}

\author{
  Eda Okur\\
  Intel Labs\\
  Anticipatory Computing Lab \\
  Hillsboro, OR, USA \\
  \texttt{eda.okur@intel.com} \\
  \And
  Sinem Aslan \\
  Intel Labs\\
  Anticipatory Computing Lab \\
  Hillsboro, OR, USA \\
  \texttt{sinem.aslan@intel.com} \\
  \And
  Nese Alyuz \\
  Intel Labs\\
  Anticipatory Computing Lab \\
  Hillsboro, OR, USA \\
  \texttt{nese.alyuz.civitci@intel.com} \\
  \And
  Asli Arslan Esme \\
  Intel Labs\\
  Anticipatory Computing Lab \\
  Hillsboro, OR, USA \\
  \texttt{asli.arslan.esme@intel.com} \\
  \And
  Ryan S. Baker \\
  University of Pennsylvania \\
  Penn Center for Learning Analytics \\
  Philadelphia, PA, USA \\
  \texttt{rybaker@upenn.edu} \\
}

\begin{document}

\maketitle

\section{Introduction}

The development of real-time affect detection models often depends upon obtaining annotated data for supervised learning by employing human experts to label the student data. One open question in annotating affective data for affect detection is whether the labelers (i.e., human experts) need to be socio-culturally similar to the students being labeled, as this impacts the cost feasibility of obtaining the labels. In this study, we investigate the following research questions: For affective state annotation, how does the socio-cultural background of human expert labelers, compared to the subjects, impact the degree of consensus and distribution of affective states obtained? Secondly, how do differences in labeler background impact the performance of affect detection models that are trained using these labels? 

\section{Methodology}
\label{method}

We employed 5 experts from the United States and 5 experts from Turkey to label the same data collected through authentic classroom pilots with students in Turkey. Using HELP \cite{ET-2017}, each group labeled 14 hours of multi-modal data collected from ten 9\textsuperscript{th} grade students in 2 sessions (40 mins each) for 3 affective states: \textit{Satisfied}, \textit{Bored}, and \textit{Confused}. We analyzed within-country and cross-country inter-rater agreements using Krippendorff’s alpha \cite{Krippendorff-2011, Siegert-2014}, where we checked all-5 and the best-3 experts (having the highest agreement) of each group. We also compared affective state distributions using majority labels obtained by each group. 

For affect detection models, we employed two modalities: (1) \textit{Appearance (Appr)}: upper-body information from the camera, (2) \textit{Context \& Performance (C\&P)}: interaction and performance logs from the online learning platform for Math. For \textit{Appr}, the raw video data are segmented into instances and time series analysis methods were utilized to extract 188 appearance features, consisting of motion and energy measures, robust statistical estimators of head velocity, and frequency domain features related to head position, pose, and facial expressions. Further details of the \textit{Appr} modality can be found in our previous study \cite{AIED-2017} where we used the
same features in this study. For \textit{C\&P}, we extracted 24 features related to time (time spent on video/questions), grade (success/failure of attempts), hints (number of hints used on questions), attempts (number of trials), and others (gender). Further details of the \textit{C\&P} features employed, which are adapted from the study \cite{Pardos-2014}, can be found in our previous study \cite{MIE-ICMI-2017}. Separate generic classifiers (Random Forests) trained using majority labels from each expert group for each modality and each activity type (\textit{Instructional} and \textit{Assessment}). Instances are sliding windows of 8-sec with 4-sec overlaps. Further details of the methodology used in this study can be found in the full version of this paper \cite{AIED-2018}.

\begin{table}[!t]
  \caption{Inter-rater agreements (Krippendorff’s Alpha) among experts from the United States (US) and Turkey (TR).}
  \label{T1}
  \centering
  \begin{tabular}{*3c}
    \toprule
     & \textbf{Human Experts} & \textbf{Krippendorff’s Alpha} \\
    \toprule
    Within-country (all-5) & US-all-5 & 0.472 \\
     & TR-all-5 & 0.585 \\
    \midrule
    Within-country (best-3) & US-best-3 & 0.564 \\
     & TR-best-3 & 0.626 \\
    \midrule
    Cross-country & Cross-all-10 & 0.379 \\
     & Cross-best-6 & 0.400 \\
    \bottomrule
  \end{tabular}
\end{table}

\begin{table}[!b]
  \caption{Affect detection classifier results (F1-scores) for separate modalities (Appr: \textit{Appearance}, C\&P: \textit{Context \& Performance}) and section types (Instr: \textit{Instructional}, Assess: \textit{Assessment}) trained using labels by experts from the United States (US) and Turkey (TR).}
  \label{T2}
  \centering
  \begin{tabular}{*6c}
    \toprule
     & & \multicolumn{2}{c}{\textbf{Labels: US}} & \multicolumn{2}{c}{\textbf{Labels: TR}} \\
     \cmidrule{3-6}
     \textbf{Section Type} & \textbf{Class} & \textbf{Appr} & \textbf{C\&P} & \textbf{Appr} & \textbf{C\&P} \\
    \toprule
    Instructional & Satisfied & 0.62 & 0.58 & 0.41 & 0.42 \\
     & Bored & \underline{0.67} & \underline{0.59} & \underline{0.86} & \underline{0.88} \\
     \cmidrule{2-6}
     & Overall & 0.65 & 0.58 & 0.77 & 0.80 \\
    \midrule
    Assessment & Satisfied & 0.59 & 0.80 & 0.43 & 0.73 \\
     & Confused & \underline{0.45} & \underline{0.63} & \underline{0.57} & \underline{0.66} \\
     \cmidrule{2-6}
     & Overall & 0.53 & 0.74 & 0.51 & 0.70 \\
    \bottomrule
  \end{tabular}
\end{table}

\begin{figure}[h]
  \centering
  \includegraphics[width=\textwidth]{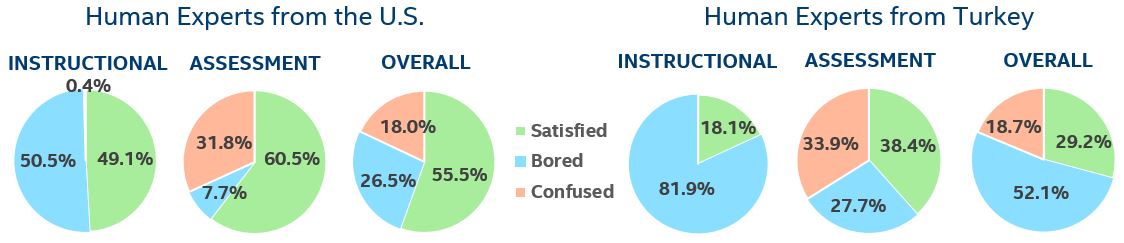}
  \caption{Affective-state distributions by the experts from US and TR}
  \label{fig:dist}
\end{figure}

\section{Experimental Results}
\label{exp_res}

The inter-rater agreements and affect detection model results are summarized in Table~\ref{T1} and Table~\ref{T2}, respectively. Students’ affective states distributions are given in Figure~\ref{fig:dist}. These results indicate that experts from Turkey obtained moderately better inter-rater agreement than the experts from the U.S. Note that even though the U.S. experts agree with each other, they agree fairly poorly with the Turkey experts. In addition, we observed important differences between the distributions of affective states provided by experts in the U.S. versus Turkey, and between the performances of the resulting real-time multi-modal affect detectors; especially for \textit{Bored} and \textit{Confused} states.

\section{Discussion and Conclusion}

Our findings suggest that there are indeed implications to using expert labelers who do not belong to the same population as the research subjects. The results in this study indicate that there could be a cultural impact in interpreting labeling ambiguities for affective states, which also has an impact on the affect detection model accuracies, especially for detecting the \textit{Bored} and \textit{Confused} states of the students. One key take-away message from this research is that cross-national or cross-cultural labelers should be vetted for inter-rater agreement very carefully \cite{BROMP-2015}.

\small

\bibliography{main}

\begin{thebibliography}{8}
\providecommand{\natexlab}[1]{#1}
\providecommand{\url}[1]{\texttt{#1}}
\expandafter\ifx\csname urlstyle\endcsname\relax
  \providecommand{\doi}[1]{doi: #1}\else
  \providecommand{\doi}{doi: \begingroup \urlstyle{rm}\Url}\fi

\bibitem[Alyuz et~al.(2017)Alyuz, Okur, Genc, Aslan, Tanriover, and
  Esme]{MIE-ICMI-2017}
N.~Alyuz, E.~Okur, U.~Genc, S.~Aslan, C.~Tanriover, and A.~A. Esme.
\newblock An unobtrusive and multimodal approach for behavioral engagement
  detection of students.
\newblock In \emph{Proceedings of the 1st ACM SIGCHI International Workshop on
  Multimodal Interaction for Education}, MIE 2017, pages 26--32, New York, NY,
  USA, 2017. ACM.
\newblock ISBN 978-1-4503-5557-5.
\newblock \doi{10.1145/3139513.3139521}.
\newblock URL \url{https://doi.acm.org/10.1145/3139513.3139521}.

\bibitem[Aslan et~al.(2017)Aslan, Mete, Okur, Oktay, Alyuz, Genc, Stanhill, and
  Esme]{ET-2017}
S.~Aslan, S.~E. Mete, E.~Okur, E.~Oktay, N.~Alyuz, U.~E. Genc, D.~Stanhill, and
  A.~A. Esme.
\newblock Human expert labeling process (help): Towards a reliable higher-order
  user state labeling process and tool to assess student engagement.
\newblock \emph{Educational Technology}, 57\penalty0 (1):\penalty0 53--59,
  2017.
\newblock ISSN 00131962.
\newblock URL \url{https://eric.ed.gov/?id=EJ1126255}.

\bibitem[Krippendorff(2011)]{Krippendorff-2011}
K.~Krippendorff.
\newblock Computing krippendorff's alpha-reliability.
\newblock \emph{Departmental Papers (ASC)}, \penalty0 (43), 2011.
\newblock URL \url{https://repository.upenn.edu/asc_papers/43}.

\bibitem[Ocumpaugh(2015)]{BROMP-2015}
J.~Ocumpaugh.
\newblock Baker rodrigo ocumpaugh monitoring protocol (bromp) 2.0 technical and
  training manual.
\newblock \emph{New York, NY and Manila, Philippines: Teachers College,
  Columbia University and Ateneo Laboratory for the Learning Sciences}, 2015.

\bibitem[Okur et~al.(2017)Okur, Alyuz, Aslan, Genc, Tanriover, and
  Arslan~Esme]{AIED-2017}
E.~Okur, N.~Alyuz, S.~Aslan, U.~Genc, C.~Tanriover, and A.~Arslan~Esme.
\newblock Behavioral engagement detection of students in the wild.
\newblock In \emph{International Conference on Artificial Intelligence in
  Education (AIED 2017)}, volume 10331 of \emph{Lecture Notes in Computer
  Science}, pages 250--261, Cham, June 2017. Springer International Publishing.
\newblock ISBN 978-3-319-61425-0.
\newblock \doi{10.1007/978-3-319-61425-0_21}.
\newblock URL \url{https://doi.org/10.1007/978-3-319-61425-0_21}.

\bibitem[Okur et~al.(2018)Okur, Aslan, Alyuz, Arslan~Esme, and
  Baker]{AIED-2018}
E.~Okur, S.~Aslan, N.~Alyuz, A.~Arslan~Esme, and R.~S. Baker.
\newblock Role of socio-cultural differences in labeling students' affective
  states.
\newblock In \emph{International Conference on Artificial Intelligence in
  Education (AIED 2018)}, volume 10947 of \emph{Lecture Notes in Computer
  Science}, pages 367--380, Cham, June 2018. Springer International Publishing.
\newblock ISBN 978-3-319-93843-1.
\newblock \doi{10.1007/978-3-319-93843-1_27}.
\newblock URL \url{https://doi.org/10.1007/978-3-319-93843-1_27}.

\bibitem[Pardos et~al.(2014)Pardos, Baker, San~Pedro, Gowda, and
  Gowda]{Pardos-2014}
Z.~A. Pardos, R.~S. Baker, M.~San~Pedro, S.~M. Gowda, and S.~M. Gowda.
\newblock Affective states and state tests: investigating how affect and
  engagement during the school year predict end-of-year learning outcomes.
\newblock \emph{Journal of Learning Analytics}, 1\penalty0 (1):\penalty0
  107--128, 2014.

\bibitem[Siegert et~al.(2014)Siegert, B{\"o}ck, and Wendemuth]{Siegert-2014}
I.~Siegert, R.~B{\"o}ck, and A.~Wendemuth.
\newblock Inter-rater reliability for emotion annotation in human--computer
  interaction: comparison and methodological improvements.
\newblock \emph{Journal on Multimodal User Interfaces}, 8\penalty0
  (1):\penalty0 17--28, Mar 2014.
\newblock ISSN 1783-8738.
\newblock \doi{10.1007/s12193-013-0129-9}.
\newblock URL \url{https://doi.org/10.1007/s12193-013-0129-9}.

\end{thebibliography}

\end{document}